\documentclass[
 reprint,
 superscriptaddress,
 amsmath,amssymb,
 aps,
]{revtex4-1}

\usepackage{amsmath}
\usepackage{graphicx}
\usepackage{dcolumn}
\usepackage{bm}
\usepackage{mathtools}
\usepackage{gensymb}
\usepackage{amssymb}
\usepackage{csvsimple}
\usepackage[capitalize]{cleveref}
\usepackage[per-mode=reciprocal]{siunitx}

\newcommand{\Conv}{
  \mathop{\scalebox{1.5}{\raisebox{-0.2ex}{$\circledast$}}
  }
}
\newcommand{\conv}{\circledast}

\begin{document}

\title{Efficient Optical Quantification of Heterogeneous Emitter Ensembles}
\author{S. Alex Breitweiser}
\affiliation{ 
Quantum Engineering Laboratory, Department of Electrical and Systems Engineering, University of Pennsylvania, 200 S. 33rd St. Philadelphia, Pennsylvania, 19104, USA
}
\affiliation{
Department of Physics and Astronomy, University of Pennsylvania, 209 S. 33rd St. Philadelphia, Pennsylvania 19104, USA
}

\author{Annemarie L. Exarhos}
\altaffiliation{Present address: Department of Physics, Lafayette College, Easton, PA 18042, USA.}
\affiliation{ 
Quantum Engineering Laboratory, Department of Electrical and Systems Engineering, University of Pennsylvania, 200 S. 33rd St. Philadelphia, Pennsylvania, 19104, USA
}

\author{Raj N. Patel}
\affiliation{ 
Quantum Engineering Laboratory, Department of Electrical and Systems Engineering, University of Pennsylvania, 200 S. 33rd St. Philadelphia, Pennsylvania, 19104, USA
}

\author{Jennifer Saouaf}
\affiliation{
Quantum Engineering Laboratory, Department of Electrical and Systems Engineering, University of Pennsylvania, 200 S. 33rd St. Philadelphia, Pennsylvania, 19104, USA
}

\author{Benjamin Porat}
\affiliation{ 
Quantum Engineering Laboratory, Department of Electrical and Systems Engineering, University of Pennsylvania, 200 S. 33rd St. Philadelphia, Pennsylvania, 19104, USA
}

\author{David A. Hopper}
\affiliation{ 
Quantum Engineering Laboratory, Department of Electrical and Systems Engineering, University of Pennsylvania, 200 S. 33rd St. Philadelphia, Pennsylvania, 19104, USA
}
\affiliation{
Department of Physics and Astronomy, University of Pennsylvania, 209 S. 33rd St. Philadelphia, Pennsylvania 19104, USA
}

\author{Lee C. Bassett}
\email{lbassett@seas.upenn.edu}
\affiliation{ 
Quantum Engineering Laboratory, Department of Electrical and Systems Engineering, University of Pennsylvania, 200 S. 33rd St. Philadelphia, Pennsylvania, 19104, USA
}

\date{\today}

\begin{abstract}
Defect-based quantum emitters in solid state materials offer a promising platform for quantum communication and sensing.
Confocal fluorescence microscopy techniques have revealed quantum emitters in a multitude of host materials.
In some materials, however, optical properties vary widely between emitters, even within the same sample.
In these cases, traditional ensemble fluorescence measurements are confounded by heterogeneity, whereas individual defect-by-defect studies are impractical.
Here, we develop a method to quantitatively and systematically analyze the properties of heterogeneous emitter ensembles using large-area photoluminescence maps.
We apply this method to study the effects of sample treatments on emitters in hexagonal boron nitride, and we find that low-energy (\SI{3}{\kilo\electronvolt}) electron irradiation creates emitters, whereas high-temperature (\SI{850}{\celsius}) annealing in an inert gas environment brightens emitters.

\end{abstract}

\maketitle

As optically addressable spin qubits, defects in solid-state materials have been used to facilitate the storage and transmission of quantum information and precisely sense temperature, strain, and electromagnetic fields at the nano-scale.\cite{Awschalom2018, Atature2018}
The most prominent of these defects, such as the nitrogen-vacancy (NV) center in diamond, act as point-source quantum emitters and have a well-understood chemical structure that can be controllably formed in high-purity host materials.
Historically, the availability of homogeneous emitter ensembles has been essential for their identification \cite{Davies1976,Loubser1977} and development for quantum applications \cite{Acosta2009, Waldermann2007}.
Using confocal microscopy, however, it is possible to screen many potential host materials for individual quantum emitters.
Indeed, quantum emitters have been found in an ever increasing number of materials, including silicon carbide, zinc oxide, gallium nitride, hexagonal boron nitride (hBN), and the transition metal dichalcogenides \cite{Aharonovich2016,Bassett2019}.
In these emerging materials platforms, not all emitters are created equal.
Studies are confounded by difficulty synthesizing the host material and controlling its purity, uncertainty about the background impurity levels, unknown chemical structure of the emitters, and often substantial heterogeneity of the emitters themselves.

Hexagonal boron nitride (hBN), a two-dimensional (2D) semiconductor with an indirect bandgap of \SI{5.955}{\electronvolt} \cite{Cassabois2016} and a rich taxonomy of defects \cite{Mcdougall2017, Weston2018}, is a prototypical example.
HBN is already an ubiquitous dielectric in van der Waals heterostructures \cite{Geim2013}, and it is emerging as a versatile platform for nanophotonics \cite{Caldwell2019}.
Recent experiments have identified point-like quantum emission at visible to near-infrared wavelengths from hBN \cite{Tran2015, Tran2016, Exarhos2017, Chejanovsky2016, Jungwirth2016, Martinez2016}.
These emitters appear to be robust to the preparation method, having been found in hBN samples of dispersed nanoflakes, exfoliated bulk crystals, and thin films grown by chemical vapor deposition, in thicknesses ranging from monolayer to bulk \cite{Toth2019}.
The emitters can be spectrally tuned by strain \cite{Grosso2017} and electric fields \cite{Noh2018}, and they can be coupled to photonic nanocavities and dielectric antennas which direct and enhance the emission \cite{Vogl2019b, Li2019}.
Some emitters in hBN also exhibit magnetically-sensitive fluorescence at room temperature, indicating the potential for coherent spin control \cite{Exarhos2019}.

However, the underlying electronic and chemical structure of emitters in hBN has remained elusive, partly due to their heterogeneous properties.
The brightness, density, polarization, and spectral distribution of emitters varies widely, both between and within samples \cite{Jungwirth2016, Tran2016, Exarhos2017}.
Electronic structure calculations have considered multiple defect candidates as possible sources of the emission \cite{Tawfik2017, Sajid2018, Abdi2018, Weston2018}, but none can fully account for the observations.
A plethora of treatments\,---\,including annealing \cite{Tran2015}, plasma \cite{Xu2018,Vogl2018} and chemical \cite{Chejanovsky2016} etching, irradiation by electrons (both low- and high-energy) \cite{Tran2016,Choi2016,Ngoc2018} and ions \cite{Chejanovsky2016}, strain engineering \cite{Proscia2018}, and ion-beam milling \cite{Ziegler2019}\,---\,have been used to create and stabilize emitters.
Analyzing the effects of these experiments has so far relied on individually cataloguing large numbers of heterogeneous emitters by hand.
Here, we present a versatile and efficient framework for quantifying the optical properties of heterogeneous emitter ensembles from large-area photoluminescence (PL) maps, and we apply it to analyze two common sample treatments in hBN: high-temperature annealing under an inert gas environment and low-energy electron beam irradiation.

\section{\label{sec:analysis}Model and Analysis}
\begin{figure}
\includegraphics{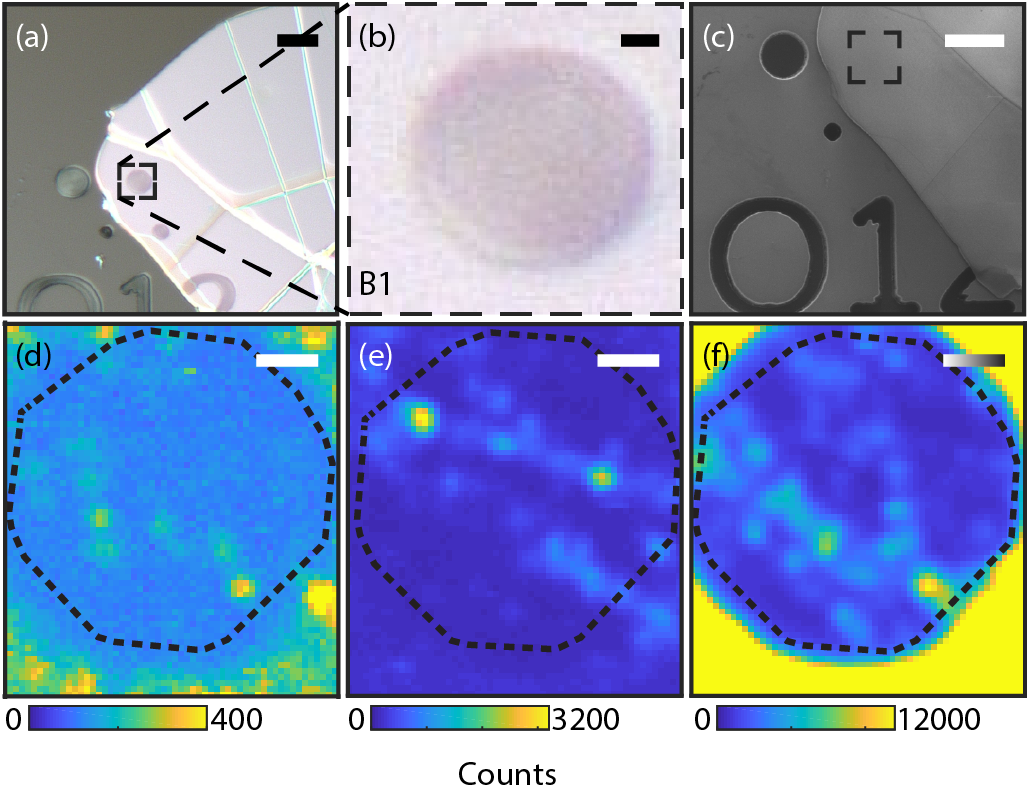}
\caption{\label{Figure1} a,b) Optical and c) electron microscope images of a flake of hBN on a patterned Si/SiO$_2$ substrate. 
The magnified window in (b) is outlined with a dashed line in (a) and shows a suspended region of hBN, labeled as Region B1 later in the text.
The same region is outlined again in the electron microscope image (c).
Profilometer measurements, available in the Supporting Information, confirm the flake is flat near this region.
The lower panels show PL maps of the same region d) before treatment, e) after electron irradiation, and f) after high temperature argon annealing.
The suspended region used in subsequent analysis is outlined with a dashed line in (d-f).
Scale bars in (a,c) represent \SI{10}{\micro\meter}, while those in (b) and (d-f) represent \SI{1}{\micro\meter}.}
\end{figure}

As an example of the type of data we wish to analyze, \cref{Figure1} shows optical and electron microscope images of an hBN flake, as well as PL maps from a suspended region taken before treatment, after electron irradiation, and after subsequent annealing.
The pre-treatment map reveals multiple emitters of similar brightness.
After undergoing electron beam irradiation, many more emitters are visible, with some now much brighter than others.
After annealing, the apparent number of emitters further increases, with the brightest emitters again much brighter than the dimmest.
While it is generally difficult to track individual emitters across treatments, the Supporting Information includes a dataset where certain emitter clusters persisted before and after irradiation, and some isolated emitters appear in the same location with similar dipole orientations before and after annealing.

Rather than attempting to identify and track every emitter in these scans, we fit the data using a model that predicts the statistical properties of heterogeneous point-source emitter ensembles.
The procedure distills the map into a distribution of pixel intensities, which is fit to produce an estimate of the density and brightness distribution of emitters present in the region. 
The model assumes that emitters appear as diffraction-limited point sources with uniform spatial distribution, and with brightness drawn from a weighted mixture of normal distributions. 
In the following analysis and discussion, we interpret these normal distributions as multiple emitter ``families,'' each characterized by a spatial density, mean brightness, and brightness variance.
We stress, however, that this is purely a phenomenological description of the observed emitter distributions; it does not necessarily reflect a classification of the underlying chemical or electronic structure of these emitters.

The model produces a probability density for the intensity of pixels in the region,
\begin{multline}
p(I|\eta_m, A_m, \sigma_m, \lambda) = \\
\Conv_{m=1}^M\left(\sum_n P_n(\eta_m)p_n(I|A_m,\sigma_m)\right)\conv\operatorname{Poiss}(I|\lambda)\,,
\end{multline}
where $m\in [1,M]$ labels each emitter family with corresponding density, $\eta_m$, average brightness, $A_m$, and brightness standard deviation, $\sigma_m$, while $\lambda$ parameterizes the brightness of the Poissonian background. 
$P_n(\eta_m)$ is the probability of having $n$ emitters of family $m$ within the region of interest, and $p_n(I|A_m,\sigma_m)$ is the probability density for pixels as a function of brightness, $I$, resulting from $n$ emitters from family $m$.
$\operatorname{Poiss}(I|\lambda)$ is the probability density resulting from a Poissonian background with average intensity $\lambda$, and $\Conv_{m=1}^M$ and $\conv$ represent convolutions. 
See the Supporting Information for a derivation of this model, along with explicit expressions for $P_n$ and $p_n$.
In general, the form of these functions depends on assumptions regarding the emitters' spatial and brightness distribution. 
We assume a uniform spatial distribution and a normal brightness distribution for each emitter family, but the model can be adapted to any spatial or brightness distribution. 
The model is probabilistic and ignores spatial information to take advantage of the statistical power of large maps; therefore it does not reveal information about individual emitters, but rather ensemble properties of the collection of emitters present in a sample.

We fit this model to the observed pixel brightness distributions using Differential Evolution \cite{Storn1997} to optimize the chi-squared statistic.
The number of families, $M$, is chosen to minimize the Akaike Information Criteria (AIC), which measures fit quality while penalizing overfitting from a high number of families.
More details are presented in the Supporting Information.

\begin{figure}
\includegraphics{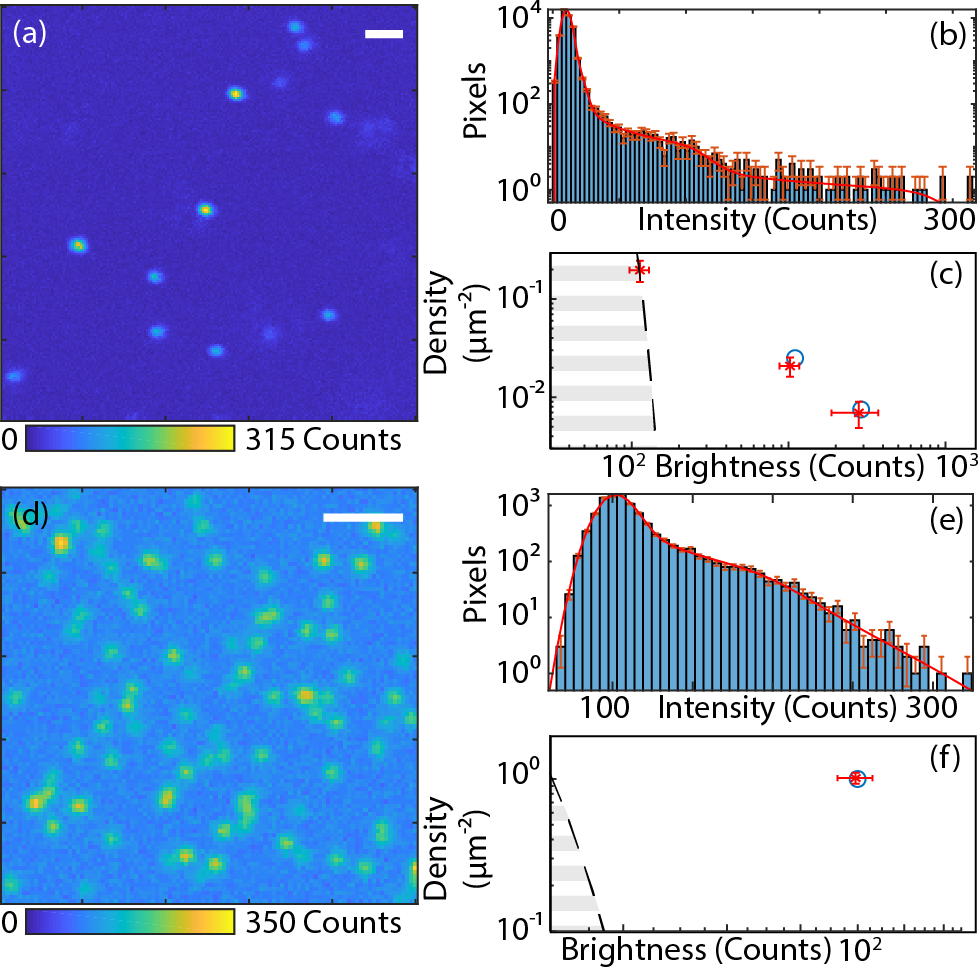}
\caption{\label{Figure2}Testing the quantitative model. 
a) PL map of NV centers in single-crystal, bulk diamond. Several NV centers, both aligned and misaligned to the excitation laser polarization, are in focus. 
b) Histogram of pixel intensities (points) from (a), together with the best fit of the quantitative model (red curve). 
c) Emitter family parameters (red crosses) corresponding to the best-fit curve in (b).
The best-fit background is represented by a dashed curve, bounding a hatched region where the model cannot resolve emitter families from noise.
Known values for the density and brightness corresponding to NV centers with different dipole orientations are shown as blue circles.
d) Simulated PL map for a single family of emitters based on parameters similar to hBN maps. 
(e-f) Corresponding pixel intensity histogram, fit, and emitter family parameter plot as in (b-c). 
The underlying simulation values for the emitter family are indicated by a blue circle.
Scale bars in (a,d) represent \SI{2}{\micro\meter}.
Error bars in (c,f) represent \SI{95}{\percent} confidence intervals.}
\end{figure}

To test this procedure, we compared the fit results from the model against the parameters of known emitter distributions. \Cref{Figure2}(a) shows a PL map of NV centers in bulk, single-crystal diamond, with a focus plane located approximately \SI{3}{\micro\meter} from the planar, (100)-oriented surface.
The laser polarization is aligned to the dominant optical excitation dipole for the NV center at the center of the map. 
There are three aligned and ten misaligned NV centers in this \SI{400}{\square\micro\meter} region, with reproducible peak intensities of $\approx$ \SI{300}{Counts} (\SI{30}{\kilo Cts \per\second}) and $\approx$ \SI{120}{Counts} (\SI{12}{\kilo Cts \per\second}), respectively. 
These peak intensities include a background of $\approx$ \SI{10}{Counts} (\SI{1}{\kilo Cts\per\second}), which appears to be uniform across the map. 
In addition, some non-point-like emission appears in \cref{Figure2}(a), which may result from out-of-focus NV centers or surface contamination.

\Cref{Figure2}(b) shows the histogram of pixel intensities from this map, as well as the result of fitting the model to this distribution.
The fitting procedure identifies three emitter families, whose density and brightness parameters are shown in \cref{Figure2}(c). 
Two of these families are within one standard uncertainty of both the density and brightness of the aligned and misaligned NV centers identified in the map, after accounting for the background.
In addition, the best-fit background of \SI{10.064 \pm 0.092}{Counts} is close to the $\approx$ \SI{10}{Counts} measured by eye from the PL map, and is represented in \cref{Figure2}(c) by a dashed curve in brightness/density space. 
For values of brightness and density below this curve, the model cannot reliably distinguish emitter families from noise.
One additional emitter family appears close to the noise floor in \cref{Figure2}(c); this may arise from weak, non-point-source PL features in the map.

The low density and reproducible brightness of NV centers in diamond make maps like \cref{Figure2}(a) easy to interpret by eye. 
A simulated dataset with less ideal conditions, similar to the hBN PL maps of \cref{Figure1}(d-f), is shown in \cref{Figure2}(d). 
Here the density of emitters is much greater, such that some emitters overlap and are indistinguishable by eye, and the brightness of emitters has a wide distribution. 
In addition, the background intensity is comparable to the brightness of emitters.
Nevertheless, the fitting procedure captures the pixel intensity distribution well, and the analysis yields a single emitter family with parameters that agree with the underlying simulation parameters, as shown in \cref{Figure2}(e,f). 
Similar analyses for multiple emitter families are presented in the Supporting Information.

\section{\label{sec:results}Results}

\begin{table}
\begin{tabular}{|l|l|l|l|l|}
\hline
Region & Thickness     & 1st Treatment    & 2nd Treatment \\
\hline
A1     & 215nm               & Low-dose e\textsuperscript{-} Irr.    & Ar Anneal        \\
\hline
A2     & 240nm               & Low-dose e\textsuperscript{-} Irr.    & Ar Anneal        \\
\hline
B1     & 390nm               & High-dose e\textsuperscript{-} Irr.   & Ar Anneal        \\
\hline
B2     & 250-350nm           & High-dose e\textsuperscript{-} Irr.   & Ar Anneal        \\
\hline
C1     & 630nm               & Indirect e\textsuperscript{-} Irr.    & Ar Anneal        \\
\hline
D1     & *                   & Ar Anneal           & Low-dose e\textsuperscript{-} Irr.\textsuperscript{\textdagger} \\
\hline
\end{tabular}

\caption{\label{tab:Regions}Summary of hBN regions and treatment sequences}
* Thickness information is not available for this region.

\textsuperscript{\textdagger} The radiation dose in this region was \SI{4e15}{e\textsuperscript{-}\per\square{\centi\meter}}.
\end{table}

\begin{figure}
\includegraphics{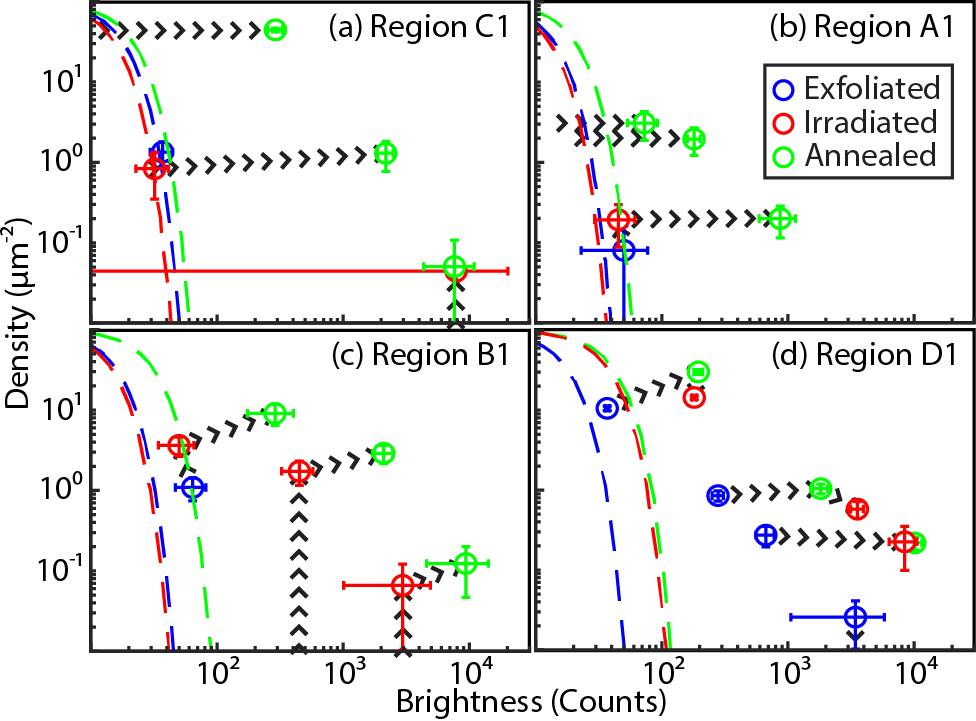}
\caption{\label{Figure3}Best-fit parameters after each treatment stage for four representative hBN regions from \cref{tab:Regions}. 
The estimated density and brightness of each emitter family is represented as a circle in blue (pre-treatment), red (post-irradiation), or green (post-annealing) based on which stage of the treatment process the flake is in. 
Note that Region D1 in (d) was annealed prior to irradiation, whereas all other regions were irradiated first. 
Arrows indicate potential evolution of these emitter families from the treatment process, based on qualitative observations. 
A noise floor, determined by the best-fit background for each map, is displayed as a dashed line and determines the lower limit for detecting emitter families.
Error bars represent 95\% confidence intervals in the best-fit parameter values.}
\end{figure}

A list of hBN regions studied in this work, as well as the treatments applied to them, is presented in \cref{tab:Regions}. 
See Methods for details of sample preparation, treatments, and data acquisition.
PL maps from each region for each stage of treatment were analyzed using our model; regions were imaged under identical conditions in each stage. 
The fitting results from four representative regions are presented in \cref{Figure3} and discussed below.
Raw PL map data, optical microscope images of the hBN samples, as well as analysis of additional regions are available in the Supporting Information.

We first consider the effect of electron irradiation.
Region C1 received no direct exposure to the electron beam, although it was present in the instrument chamber to measure effects of the ambient chamber conditions. 
As shown in \cref{Figure3}(a), this region saw a small decrease in the density of pre-existing emitters, although it was within the uncertainty of the fits.
Some decrease is expected due to photobleaching from successive scans.
In addition, the appearance of a single bright emitter resulted in the detection of a new family with high brightness and low density; we tentatively attribute this isolated event to stray ions accelerated in the column.
Regions A1 and B1 received low and high doses of direct electron irradiation, respectively. Region A1 saw a small increase in the density of its pre-existing emitter family, although the increase was within the fit uncertainty; see \cref{Figure3}(b). 
The high-dose region B1 considered in \cref{Figure3}(c) is the same one presented in \cref{Figure1}; it showed a significant increase in the density of its pre-existing emitter family, along with the appearance of two new families of higher brightness.

Regions A1, B1, and C1 were all annealed after irradiation, under the same conditions as described in Methods. 
Looking at the fit results for region B1, there are three emitter families both before and after annealing, with approximately the same densities but systematically higher brightnesses. 
The simplest interpretation of these results is that each family became brighter without a significant change in density. 
Analysis of regions A1 and C1 also uncovered post-annealing families with similar density to pre-existing families but with higher brightness, consistent with this interpretation.
However, these regions also contained new dim, dense emitter families after annealing. 
Considering the relative change in brightness for pre-existing emitter families, we propose that these emitters existed before annealing, but were below the noise floor\,---\,only becoming bright enough to be captured by the model after annealing. 
No new families are detected after annealing for the high-dose Region B1. 
However, the dimmest family saw a density increase at the edge of statistical significance, and the background showed a large increase not seen in the low-dose or non-irradiated region.
Both of these features could be indicative of a dim, dense family that cannot adequately be resolved in the data.
In \cref{Figure3}, we indicate the possible evolution of emitter families under annealing \textit{via} dashed lines.

Region D1 was first annealed, followed by a low dose of irradiation; its analysis is shown in \Cref{Figure3}(d). 
The pre-treatment analysis detects four families of emitters already present in the region; we attribute the greater number of pre-treatment emitter families here to alternate sample preparation, as this flake was exfoliated from a different bulk crystal and underwent a post-exfoliation O$_2$ plasma clean.
Recent studies have shown plasma treatments may create new emitters \cite{Xu2018, Vogl2018}.
After annealing, the brightest family, which consisted of a single emitter, disappeared, which we attribute to photobleaching.
The other three families can be seen to increase in brightness again. 
Similar to region B1, the dimmest emitter family also saw a slight increase in density, and the background saw a large increase. 
After irradiation, the emitter parameters did not exhibit a significant change.
This is consistent with the results for Region A1, which also received a low irradiation dose; the expected increase in emitter density is small compared to the large densities already present in Region D1 after annealing.

In order to compare results from different regions, and to generate a more rigorous statistical understanding of the treatment effects, we consider the full emitter brightness distribution extracted from the analysis of each region.
The brightness distribution of emitters is obtained by summing the individual contributions of each family weighted by their densities, $\sum_m \eta_m \mathcal{N}(I|A_m,\sigma_m^2)$, where $\mathcal{N}(I|A,\sigma^2)$ is a normal distribution on $I$ with mean $A$ and variance $\sigma^2$.
However, due to the wide range of brightnesses observed, the results are best shown on a logarithmic brightness scale.
Thus, we present the results as a log-space probability density,
\begin{equation}
\Lambda(I) = I*\sum_m \eta_m \mathcal{N}(I|A_m,\sigma_m^2)\,,
\end{equation}
where the additional factor of $I$ accounts for the logarithmic spacing of brightnesses, correcting the visual weight of the plotted distribution.

\begin{figure}
\includegraphics{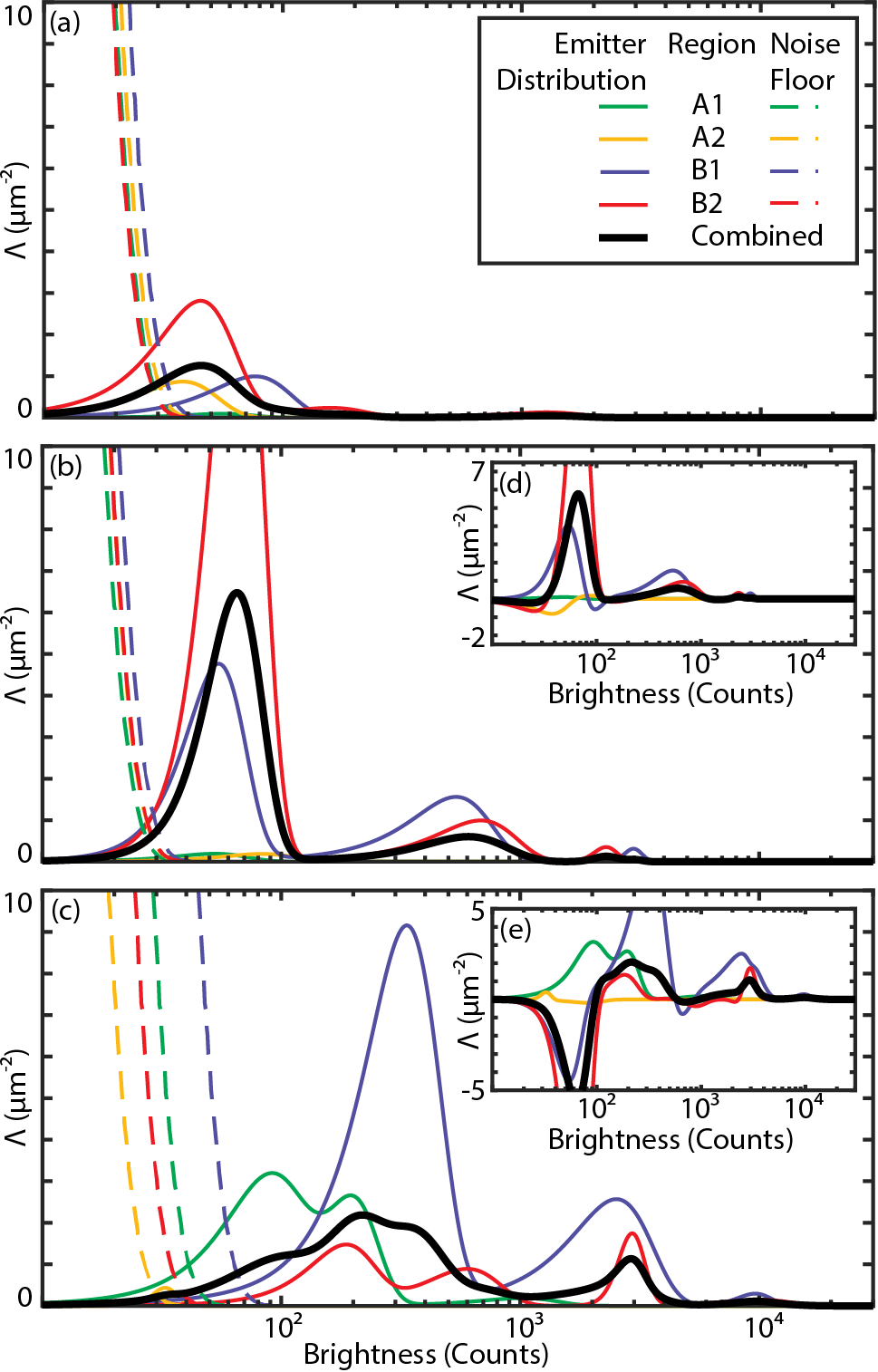}
\caption{\label{Figure4} Emitter brightness distributions for regions which received direct irradiation prior to annealing; distributions for each region a) before treatment, b) after irradiation, and c) after annealing are plotted as thin colored curves for each region as indicated in the legend, with the combined distribution for all regions shown as a thick black curve. Dashed curves indicate the background noise floor for each region.
Inset panels (d) and (e) show the changes in the distributions from pre-treatment to post-irradiation and post-irradiation to post-annealing, respectively. 
Note that, since the brightness is shown on a logarithmic scale, the log-space probability density is shown; refer to the main text for details.
}
\end{figure}

\Cref{Figure4} shows this distribution for regions which received irradiation followed by annealing.
In the pre-treatment distributions of \cref{Figure4}(a), we observe a localized peak around \SI{50}{Counts} ($\approx$ \SI{400}{Cts\per\second}), with very few emitters brighter than \SI{1000}{Counts} ($\approx$ \SI{8}{\kilo Cts\per\second}). 
Any emitters with brightness below $\approx$ \SI{30}{Counts} cannot be resolved from the background noise (indicated by dashed lines). 
\Cref{Figure4}(b) shows the same regions after irradiation. 
We observe a much larger peak around \SI{70}{Counts}, as well as new peaks appearing with higher brightness.
Finally, the post-annealing distributions shown in \cref{Figure4}(c) are much broader, with emitters found from the noise floor of $\approx$ \SI{30}{Counts} to $\approx$ \SI{e4}{Counts} ($\approx$ \SI{800}{\kilo Cts \per\second}).

To better visualize the treatment effects, \cref{Figure4}(d,e) present the difference between the distributions before and after irradiation and annealing, respectively. 
We observe from \cref{Figure4}(d) that irradiation produced an almost uniform increase in the density of emitters, and regions B1 and B2, which received larger doses, saw larger density increases than regions A1 and A2. 
Annealing showed a qualitatively different effect, with densities decreasing at lower brightness and increasing at higher brightness in \cref{Figure4}(e). 
There is a small overall increase in density; this, as well as the overall broadening of the distribution, can tentatively be attributed to the dim, dense emitters that might be below the noise floor in pre-annealing fits.

\section{\label{sec:discussion}Discussion}
\subsection{Irradiation}
In agreement with qualitative observations of images such as those in \cref{Figure1} and with the observations of earlier studies \cite{Choi2016, Tran2016, Vogl2019a}, our quantitative analysis shows that low-energy electron irradiation increases the emitter density, creating emitters of low-to-intermediate brightness.
However, the underlying mechanism for emitter creation due to electron-beam irradiation remains unknown.
Using a similar irradiation procedure, Vogl \textit{et al}. found emitters are created almost uniformly at all depths in samples thinner than the stopping range of the electrons, which is on the order of a micron for the \SI{3}{\kilo\electronvolt} energy of our electron beam \cite{Vogl2019a}.
Combined with our observation that the number of emitters created increases with electron dose, this argues for a creation mechanism based on interactions between the electrons and hBN, rather than accidental implantation of impurity ions in the SEM chamber.
Ions at this energy would only travel a few nanometers into the hBN, and they would appear independently of beam exposure.
However, the electron energies used here are much lower than the minimum knock-on energy for creating monovacancies in hBN \cite{Kotakoski2010}, corresponding to an accelerating voltage of \SI{80}{\kilo\volt}.
The electrons might have sufficient energy to perturb the placement of interstitial atoms, or to cause reconstruction of edges or extended defects.
Alternatively, charge trapping may cause emitter activation, as suggested in previous studies \cite{Proscia2018, Shotan2016}.
Irradiation could reconfigure the charge state of existing defects, converting them into a fluorescent configuration, or the reconfiguration of nearby charge traps could alter the (non)radiative relaxation pathways relevant for the visible PL.

\subsection{Annealing}
Similarly, the role of annealing on hBN's quantum emission is poorly understood.
Whereas qualitative assessment of PL maps like those in \cref{Figure1} gives the impression that annealing creates emitters, our quantitative analysis indicates that the primary ensemble effect of annealing is to brighten existing ensembles without significantly changing the density of emitters.
The brightness increase by annealing is around one order of magnitude regardless of the emitters' original brightness\,---\,a surprising result given that the emitter brightnesses span several orders of magnitude. 
Some regions also saw the appearance of dim, dense emitter families after annealing; even for regions where such families were not detected, a larger increase in the background intensity and in the density of pre-existing families point to the possibility of dim emitters that were subsumed into the background and excluded from our analysis. 
Assuming these dim emitters were also brightened by annealing, it is possible they were present before annealing, but not detected because they were below the noise floor.

The brightness enhancement could be explained by an increase in the emitters' quantum efficiency.
Potentially, annealing affects the concentration of other, non-emissive defects in the sample, which modify non-radiative decay pathways for the emissive defects.
This interpretation is supported by the varied quantum efficiencies for hBN emitters reported in the literature, which range from 6\% \cite{Li2019} to 87\% \cite{Nikolay2019}. 
This could also explain why the brightness increase is consistently around an order of magnitude, regardless of the initial brightness of the emitters. 

While a systematic increase in brightness due to annealing is the simplest interpretation of our observations, we cannot rule out all other potential effects.
Studies using rapid thermal annealing rather than a tube furnace saw an increase in zero phonon line intensity, and noted that longer annealing times led to spatial diffusion of emitters \cite{Vogl2018}.
Such diffusion, combined with the increased brightness of emitters, could give the appearance of bright emitters being created simultaneously with dim emitters being destroyed.
The mobility of point defects or impurities in the lattice during annealing might explain these effects, either by moving the underlying defects or otherwise modifying the emitters' chemical structure.
Of the single-atom vacancies and interstitial defects, only boron vacancies are expected to become mobile around \SI{800}{\celsius}, with nitrogen vacancies requiring temperatures in excess of \SI{1500}{\celsius} and interstitial defects becoming mobile near room temperature \cite{Weston2018}.
This framework presented here could be used to study the temperature dependence of annealing effects, for comparison with theoretical calculations of the onset of defect mobility for different species.

\section{Conclusion}
We presented a method to efficiently assess the optical properties of statistically large heterogeneous quantum-emitter ensembles.
Tracking systematic variations between samples or between treatments offers quantitative insight into the mechanisms at play.
In the case of hBN, 
electron irradiation provides an accessible and controllable method for creating emitters in otherwise dark samples.
For samples with dim emitters, annealing may provide a way to brighten emitters.

While this study focused on the brightness and density of emitters, the model can also be expanded to capture other properties of quantum emitters, such as their dipole orientation and spectral distribution.
Previous work studied the alignment of emitter dipole orientations to the crystallographic axes of hBN \cite{Exarhos2017}.
By extending our model to include polarized emitter families and comparing to polarization resolved data, we could leverage the statistical power of much larger emitter ensembles to study the distribution of dipole orientations.
A similar extension of our model could account for the emitters' spectra; including spectrally resolved data could reveal phenomena such as zero phonon line clustering, which has been observed in multiple recent studies \cite{Mendelson2019, Comtet2019, Stern2019}.
By adapting the underlying spatial probability distribution functions, the model can be further extended to account for emitters clustering near edges or other extended defects.

This methodology can be applied to any material hosting point-source emitters, including dispersed nanoparticles and fluorescent molecules.
Advances in fluorescence imaging have enabled the efficient acquisition of emission maps from large regions of both two- and three-dimensional samples.
Such wide-field techniques have been used to study the role of annealing temperatures on NV center quenching and formation \cite{Chakravarthi2019}, as well as to study the spectral and temporal properties of emitters in hBN \cite{Stern2019, Comtet2019}.
However, these studies relied on algorithms designed to identify and track individual emitters, which will fail at high emitter densities and low emitter brightnesses.
Combining these wide-field imaging techniques with the approach presented here paves the way to efficiently and accurately screen large ensembles of heterogeneous emitters, an important step for the identification and study of new platforms for defect-based quantum technologies.

\section{\label{sec:methods}Methods}

Flakes of hBN were exfoliated from bulk single crystals (HQ Graphene) onto patterned Si wafers with a \SI{90}{\nano\meter} layer of thermal SiO$_2$ on top.
The flakes hosting regions A1-A2, B1-B2, C1-C3, and E1 were first exfoliated onto a polydimethylsiloxane (PDMS) stamp, and then transferred onto the silicon substrate at a temperature of \SI{50}{\celsius}.
The flakes hosting regions D1-D3 were prepared following the method in Refs.~\citenum{Exarhos2017} and \citenum{Exarhos2019}, and further underwent an O$_2$ plasma clean in an oxygen barrel asher (Anatech SCE 108). 
Flake thicknesses were measured using a stylus profilometer, and ranged from \textless \SI{100}{\nano\meter} to \textgreater \SI{600}{\nano\meter}, with most flakes falling between \num{200} and \SI{400}{\nano\meter}.
Of particular interest are regions which are suspended over holes etched into the substrate (\textgreater \SI{6}{\micro\meter} deep), where emitters show a greater contrast with the background and better isolation from substrate-dependent effects.

Suspended regions were identified using an optical microscope and alternately exposed to a 3keV electron beam (FEI Strata DB235 FIB SEM) and annealed in a tube furnace under flowing Argon gas.
The electron beam was rastered over a known area, with the dosage calculated from the area size and the approximate beam current.
Low-dose irradiated regions, with the exception of D1, received fluences on the order of \SI{2e16}{e\textsuperscript{-}\per\square{\centi\meter}}, while high dose regions received approximately \SI{2e17}{e\textsuperscript{-}\per\square{\centi\meter}}.
Region D1 received a dose of \SI{4e15}{e\textsuperscript{-}\per\square{\centi\meter}}.
Calculated fluences for all regions are available in the Supporting Information.
The annealing ramp rate was set to \SI{10}{\celsius\per\minute}, leading to a heat-up period of $\approx$ \SI{1.5}{\hour} to reach \SI{850}{\celsius}.
Once \SI{850}{\celsius} was reached, the temperature was maintained for 30 minutes, after which the sample was allowed to cool back to room temperature over the course of several hours.
Argon gas was flowed from before heating until after cooling the sample to ensure complete evacuation of other gasses from the chamber while the sample temperature was elevated.

Samples were mounted in a home-built confocal fluorescence microscope, where PL was stimulated with a \SI{592}{\nano\meter} continuous wave laser (MPB Communications, VFL-592) and collected between 650 and \SI{900}{\nano\meter}.
For this study, the pre-objective power was fixed to $\approx$ \SI{500}{\micro\watt} and the laser polarization was rotated using a half waveplate (Newport 10RP12-16) and corrected for birefringence.
PL maps were recorded for each region with multiple laser polarizations and registered, then added together to create a polarization-independent PL map.
Maps were acquired for each region at each stage of the treatment process and compared to a point emitter model, as described in the main text, to determine the underlying microscopic parameters.

For the NV-center reference scan in \cref{Figure2}(a), an electronics grade type IIa synthetic diamond from Element Six was irradiated with \SI{2}{\mega\electronvolt} electrons at a fluence of \SI{e14}{ e\textsuperscript{-}\per\square{\centi\meter}} and then annealed in forming gas at \SI{800}{\celsius} for 1 hour.
The diamond sample was mounted in another home-built confocal fluorescence microscope.
A \SI{532}{\nano\meter} continuous wave laser (Gem 532, Laser Quantum) was used to optically excite the NV centers and fluorescence was collected with a \SI{650}{\nano\meter} long-pass filter.
Polarization was similarly varied using a half wave plate and the pre-objective laser power was set to \SI{500}{\micro\watt}.

A layer of NV centers was found approximately \SI{3}{\micro\meter} from the diamond surface, and the confocal depth was set to focus on the NV at the center of the scan in \cref{Figure2}(a). The laser polarization was set to maximize the PL from the central NV center. Due to the geometry of NV centers beneath a (100) diamond surface, all other NV centers in the sample are either aligned or misaligned to the excitation axis by the same angle.

\section{Acknowledgements}
The authors thank R. Grote for assistance with electron-beam irradiation.
This work was supported by the National Science Foundation (DMR-1922278) and the Army Research Office (W911NF-15-1-0589). We gratefully acknowledge use of facilities and instrumentation at the Singh Center for Nanotechnology supported by the National Science Foundation through the National Nanotechnology Coordinated Infrastructure Program (NNCI-1542153) and Penn's Materials Research Science and Engineering Center (DMR-1720530).


\bibliography{ms}
\end{document}


\title{Supporting Information: Efficient Optical Quantification of Heterogeneous Emitter Ensembles}

\date{\today}

\author{S. Alex Breitweiser}
\affiliation{ 
Quantum Engineering Laboratory, Department of Electrical and Systems Engineering, University of Pennsylvania, 200 S. 33rd St. Philadelphia, Pennsylvania, 19104, USA
}
\affiliation{
Department of Physics and Astronomy, University of Pennsylvania, 209 S. 33rd St. Philadelphia, Pennsylvania 19104, USA
}

\author{Annemarie L. Exarhos}
\altaffiliation{Present address: Department of Physics, Lafayette College, Easton, PA 18042, USA.}
\affiliation{ 
Quantum Engineering Laboratory, Department of Electrical and Systems Engineering, University of Pennsylvania, 200 S. 33rd St. Philadelphia, Pennsylvania, 19104, USA
}

\author{Raj N. Patel}
\affiliation{ 
Quantum Engineering Laboratory, Department of Electrical and Systems Engineering, University of Pennsylvania, 200 S. 33rd St. Philadelphia, Pennsylvania, 19104, USA
}

\author{Jennifer Saouaf}
\affiliation{
Quantum Engineering Laboratory, Department of Electrical and Systems Engineering, University of Pennsylvania, 200 S. 33rd St. Philadelphia, Pennsylvania, 19104, USA
}

\author{Benjamin Porat}
\affiliation{ 
Quantum Engineering Laboratory, Department of Electrical and Systems Engineering, University of Pennsylvania, 200 S. 33rd St. Philadelphia, Pennsylvania, 19104, USA
}

\author{David A. Hopper}
\affiliation{ 
Quantum Engineering Laboratory, Department of Electrical and Systems Engineering, University of Pennsylvania, 200 S. 33rd St. Philadelphia, Pennsylvania, 19104, USA
}
\affiliation{
Department of Physics and Astronomy, University of Pennsylvania, 209 S. 33rd St. Philadelphia, Pennsylvania 19104, USA
}

\author{Lee C. Bassett}
\email{lbassett@seas.upenn.edu}
\affiliation{ 
Quantum Engineering Laboratory, Department of Electrical and Systems Engineering, University of Pennsylvania, 200 S. 33rd St. Philadelphia, Pennsylvania, 19104, USA
}

\maketitle

\onecolumngrid

\section{Height Measurements}
\begin{figure}[H]
\centering
\includegraphics[width=0.75\columnwidth]{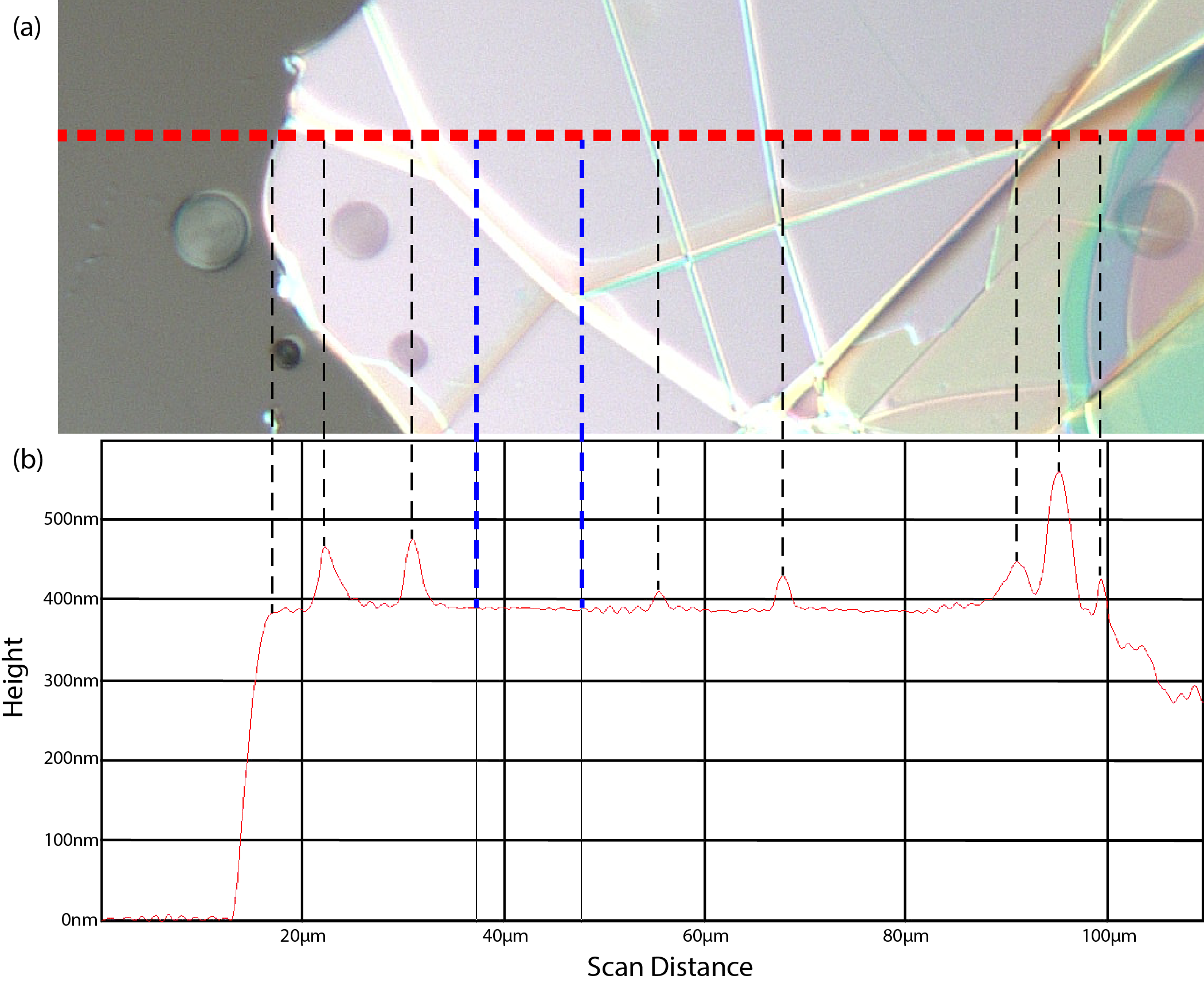}
\caption{\label{O12_profilometer}a)Optical image and b) profilometer height measurement of the flake hosting regions B1 and B2. Features in the height measurement are correlated with the optical image using dashed lines. From this, the approximate location of the scan is indicated using a dashed red line. Between the two dashed blue lines, the average height is \SI{388.1}{\nano\meter}.}
\end{figure}

\Cref{O12_profilometer} shows a profilometer height measurement of the flake hosting regions B1 and B2, along with an optical image of the flake. By correlating features in the height scan with features in the optical image (dashed black lines), we approximate the location of the profilometer scan (dashed red line). With the exception of optically visible ridges, the flake is flat in the region hosting region B1. The height is approximately \SI{388.1}{\nano\meter}, based on the average height between the two dashed blue lines.

\section{\label{sec:tracking}Tracking Emitters}
\begin{figure}[H]
\centering
\includegraphics[width=0.75\columnwidth]{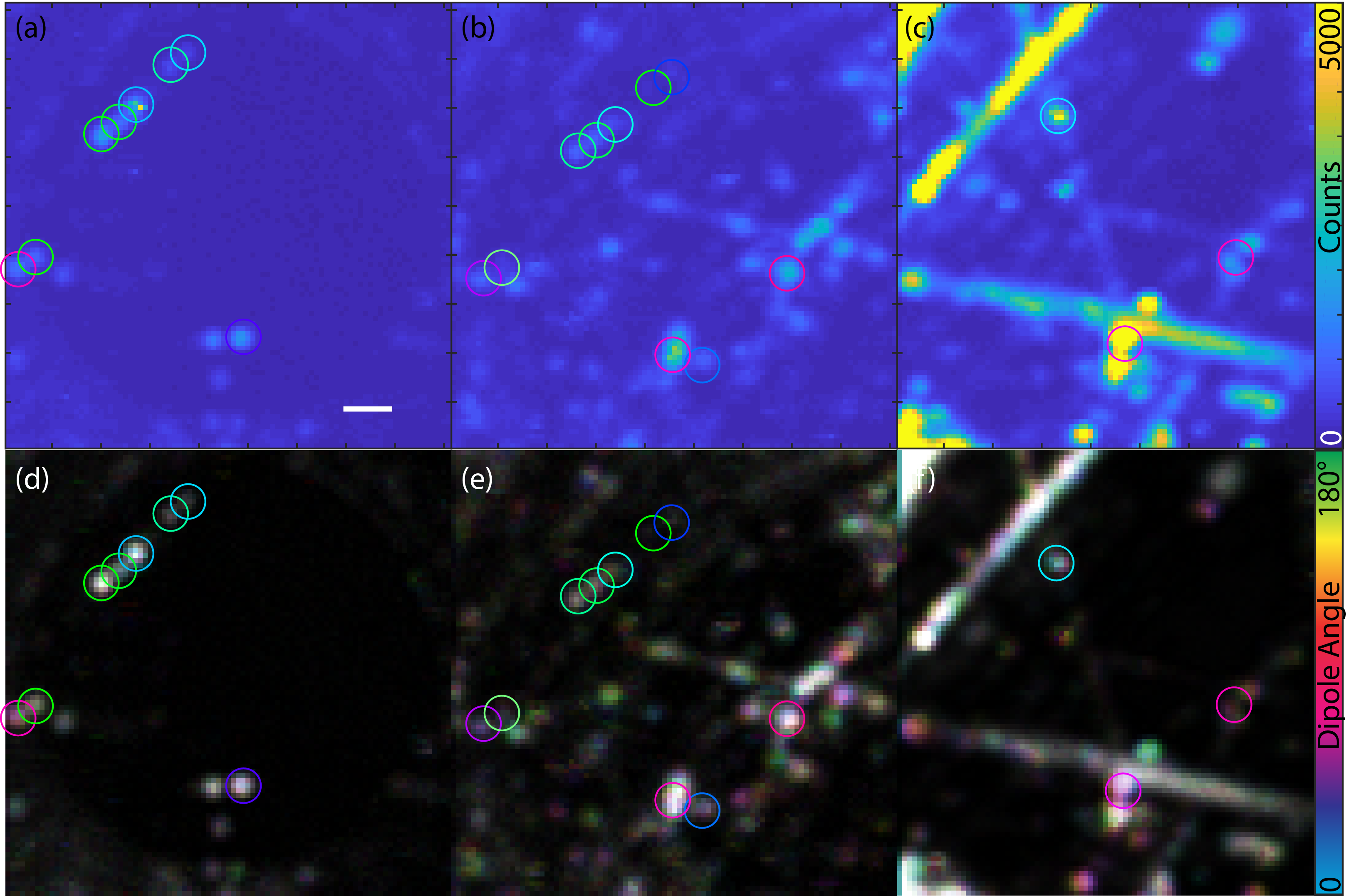}
\caption{\label{O13_tracking}Some emitters in Region B2 that seem to persist between treatments. (a-c) PL Intensity maps for Region B2 before treatment, after electron irradiation, and after annealing. (d-f) Polarization resolved PL, shown using a false color algorithm. Circles with fixed relative positions show emitters which may have persisted between treatments. The color of the circle represents the approximate polarization of the emitter. The white scale bar in (a) represents \SI{1}{\micro\meter}.}
\end{figure}


\Cref{O13_tracking} shows emitters which seem to persist between treatments in region B2.
Eight emitters are identified before and after annealing, one of which is still visible after annealing.
A combination of the consistent spatial layouts and similar dipole orientations support these identifications.
The large-scale preservation of many emitters in similar supports our conclusion that irradiation does not significantly alter existing emitters, instead primarily creating new emitters. 
Three emitters identified after irradiation are also visible after annealing, including two which seem to have been created during irradiation.
While not all emitters persist in the same location after annealing, two that did saw a significant increase in brightness, supporting our interpretation that annealing primarily increases the brightness of emitters.

\section{\label{sec:models}Emitter Model}
We assume the emitters are point emitters, with Gaussian broadening from the laser beam width. Therefore, the total photoluminescence at a point is described by
\begin{equation}
I(r) = \sum_iA_ie^{\frac{-(r-r_i)^2}{2\sigma^2}}\,,
\end{equation}
where $\sum_i$ is a sum over all the emitters, $A_i$ describes an individual emitter's photoluminescence upon direct irradiation, $r-r_i$ is the distance from the emitter to the sampled point, and $\sigma$ is the gaussian width of the beam.

For the scans used in this study this is a good approximation for the isolated emitters. In hBN scans, $\sigma$ varies from \SI{120}{\nano\meter} for a well-focused spot to \SI{200}{\nano\meter} for the worst focused spots. This variation is due to the change in focus over large area scans. We take $\sigma = \SI{150}{\nano\meter}$ as a zeroth order approximation to the data, which is representative of the average over multiple isolated emitters in multiple scans. For the NV centers in planar diamond, $\sigma$ varied from \SI{195}{\nano\meter} to \SI{220}{\nano\meter}, and $\sigma = \SI{210}{\nano\meter}$ was taken as the best approximation.

\subsection{Single emitter}
We start by analyzing the case of a single emitter randomly placed in the sample. For simplicity, we assume the sample is a square, with the understanding that as the sample size increases the large-scale geometry become insignificant due to the exponential decrease in response to far away emitters.

\subsubsection{Probability of emitter position}
To calculate the probability distribution of pixel intensities, we start with the probability distribution for separation between two points in a square of side length $a$, given by \cite{Philip2007}
\begin{equation}
p(d) = 2d\left(-4\frac{d}{a^3} + \frac{\pi}{a^2} + \frac{d^2}{a^4}\right) \quad \textrm{for} \quad 0<d<a \,,
\end{equation}
where we implicitly assume the probability density to be zero elsewhere. There is, of course, a small probability for $a < d < \sqrt{2}a$, but we ignore this as it creates only an exponentially small correction to I.

We can then use the formula
\begin{equation}
p(I) = p(d(I))\frac{d}{dI}(d)
\end{equation}
with the gaussian
\begin{equation}
I(d) = Ae^{\frac{-d^2}{2\sigma^2}}
\end{equation}
to get the photoluminescence probability density from a single emitter of known brightness
\begin{equation}
p(I|A) = \frac{2\sigma^2}{Ia^2}\left(\pi - \frac{4\sqrt{2}\sigma\sqrt{log(-\frac{I}{A}})}{a} + \frac{2\sigma^2log(-\frac{I}{A})}{a^2}\right)
	\quad \textrm{for} \quad e^{\frac{-a^2}{2\sigma^2}}<\frac{I}{A}<1\,.
\end{equation}
Because we have ignored the slight probability of $a < d < \sqrt{2}a$, this is not quite normalized, having a total probability of $\frac{-13 + 6 \pi}{6} \approx 0.975$. So we impose
\begin{equation}
p(I|A) = \left[1 - \int_{Ae^{\frac{-a^2}{2\sigma^2}}}^\infty p(I|A)dI\right]\delta(I) \quad \textrm{for} \quad \frac{I}{A}<e^{\frac{-a^2}{2\sigma^2}}\,,
\end{equation}
ignoring the fine structure of these very small intensities (which becomes exact in the large $\frac{a}{\sigma}$ limit). Note that this normalization factor will change once we quantize our model below, but the same formula is used to ensure normalization.

\subsubsection{Averaging over brightness distribution}
In the previous section, we fixed the brightness of a single emitter to $A$, but the inhomogeneous emitters observed in the samples studied require a more general brightness distribution. We begin by assuming the brightness of emitters is uniformly distributed between 0 and some maximum brightness, $A_{max}$. Then, we have the averaged single-emitter probability density
\begin{equation}
p(I|[0,A_{max}]) =\frac{ \int_0^{A_{max}}p(I|A)dA}{A_{max}}\,,
\end{equation}
where $p(I|A)$ is the intensity distribution given $A$, as specified above. This yields an analytical solution, according to Mathematica, of
\begin{equation}
p(I|[0,A_{max}]) = \frac{2\sigma^2}{IA_{max}a^2} \times
\left[\pi(A_{max}-I) + 2\frac{\sigma^2}{a^2}((I-A_{max}) + A_{max}log(\frac{A_{max}}{I}))\right]
\textrm{for} \quad e^{\frac{-a^2}{2\sigma^2}}<\frac{I}{A_{max}}<1\,.
\end{equation}
We can then easily convert this to a uniform distribution between a minimum and maximum intensity by
\begin{equation}
p(I|[A_{min},A_{max}]) =
\frac{A_{max}p(I|[0,A_{max}]) - A_{min}p(I|[0,A_{min}])}{A_{max} - A_{min}}\,.
\end{equation}

\begin{figure}[H]
\centering
\includegraphics[width=0.5\columnwidth]{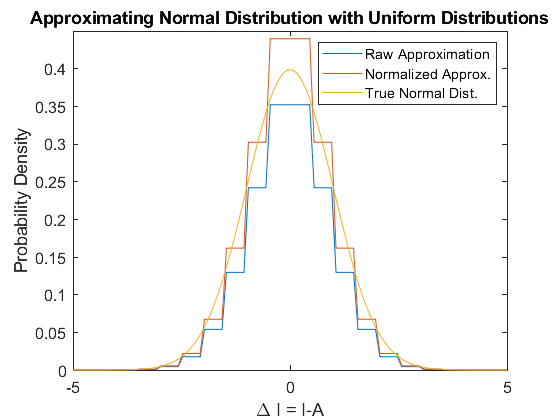}
\caption{\label{gaussian_approx}A simple approximation of a Normal Distribution by flat uniform distributions. The normalized approximation is used to optimize the numerical efficiency of the model.}
\end{figure}
With a fast numerical formula to generate flat distribution probability densities, we can approximate normal distributions to arbitrarily accuracy as a weighted sum of uniform distributions between these points. This can be made precise by increasing the number of uniform distributions, at the cost of additional computational overhead. Since the normal distribution is concave, this is a strictly lower approximation, so to ensure densities are not affected by this we also normalize the approximation. For brevity and clarity, we present this graphically, in \cref{gaussian_approx}, rather than algebraically. This gives us our final probability density for a normally distributed family of emitters as a weighted sum of uniform distributions
\begin{equation}
p(I|N(A,\sigma)) \approx \sum_iw_ip(I|[A-s_i, a+s_i])\,.
\end{equation}

\subsection{Multiple emitters}
\subsubsection{PDF for n emitters}
Denoting the probability density we calculated for a single emitter as $p_1(I)$, we can calculate the brightness probability density for two emitters as the self-convolution
\begin{equation}
p_2(I) = (p_1\conv p_1)(I)\,,
\end{equation}
where we have assumed the emitters are drawn from the same brightness distribution, but are otherwise independent in density and brightness. Similarly for $n$ emitters,
\begin{equation}
p_n(I) = (p_{n-1}\conv p_1)(I) = \underbrace{(p_1\conv...\conv p_1)}_\text{$n$-times}(I)\,.
\end{equation}
Note that, using recursive methods, this convolution can be done in $O(log(n))$ time, making it computationally feasible for even a large number of defects.
For consistency and obvious physical reasons, we define the zero emitter distribution
\begin{equation}
p_0(I) = \delta(I)\,.
\end{equation}

\subsubsection{Averaging over number of emitters}
If the probability of having $n$ emitters is labeled $P_n$, the total probability distribution is then
\begin{equation}
p(I|P_n) = \sum_{n=0}^\infty P_np_n(I)\,.
\end{equation}
We assume a large area, $a^2$, and therefore a number of emitters $N = a^2\eta$, where $\eta$ is the density of emitters. To make this a continuous parameter, we set
\begin{equation}
P_{\floor{N}} = (\ceil{N} - N) \textrm{,} \quad P_{\ceil{N}} = (N - \floor{N})\textrm{,} 
\quad \textrm{and} \quad P_n = 0 \quad \textrm{otherwise}\,,
\end{equation}
where $\floor{N}$ and $\ceil{N}$ are the floor and ceiling functions, respectively.
Note that, since the number of emitters in a scan is fixed, we do not use a Poisson distribution here.

\subsubsection{Multiple families of emitters}
To handle $m$ families of emitters, each with their own photoluminescence probability distribution $p^i(I)$ with $i=1...m$, we convolve the probability distributions together:
\begin{equation}
p_{emitters}(I) = (p^1\conv...\conv p^m)(I)\,.
\end{equation}

\subsection{Background Intensity}
A Poisson background distribution was used to fit the background, giving a single background parameter, $\lambda$, and thus
\begin{equation}
p(I) =
\left(p_{emitters} \conv p_{background}(\lambda)\right)(I)\,.
\end{equation}
Since the intensities in this study rise well above unity, we approximate a Poisson distribution by a normal distribution with equal mean and variance.

\subsection{Parameterizing the model}
With the above discussion informing our choices, we therefore parameterize the model in terms of $m$ underlying densities and brightness distributions with a single background parameter

\begin{equation}
p(I|\eta_m, A_m, \sigma_m, \lambda) =
\Conv_{m=1}^M\left(\sum_n P_n(\eta_m)p_n(I|A_m,\sigma_m)\right)\conv\operatorname{Poiss}(I|\lambda)
\end{equation}

\subsection{Quantizing the Model}
Throughout this analysis, we have taken both $I$ and $A$ to be continuous variables, allowed to take on any positive real value. However, since the data is recorded in photon counts, only integer values are possible for $I$. To account for both of this, we sampled the probability distributions for this model in quantized steps, replacing, e.g., integrals with weighted sums. Ideally sampling would occur at every whole number to capture the fundamental quantization of $I$; however, for scans with very bright emitters, it was necessary to increase this quantization further to improve the computational time necessary to optimize the parameters, as described below. In every case, this quantization was well below the counts and each of the model parameters, meaning it should not significantly affect the accuracy of the model.

Since the underlying probability distributions are quantized, we also similarly quantize the model parameters, which further reduces the parameter space and allows for better parameter estimation.

\subsection{Parameter Estimation}
First, a region of interest is defined - which may be used to mask out regions of clearly different properties (supported vs unsupported, avoiding extended defects, etc).
From this a distribution of pixel intensities is defined, grouped into an appropriate number of bins.
We then set a number of emitter families, $n$, and run an optimization algorithm to determine the best model parameters to approximate the data distribution,
\begin{equation}
\mathbf{x} = [\eta_1, A_1, \sigma_1, ... \eta_n, A_n, \sigma_n, \lambda]\,.
\end{equation}
Our optimization target function is the Neyman modified chi-squared parameter,
\begin{equation}
\tilde{\chi}^2 \equiv \sum_i\frac{(p_i(\mathbf{x}) - m_i)^2}{max(m_i,W)}\,,
\end{equation}
where the sum is taken over the data bins, $p_i(\mathbf{x})$ is the predicted number of pixels in the bin from the above model distribution given the parameter vector $\mathbf{x}$, and $m_i$ is the measured number of data pixels in that bin. $W$ is a parameter which accounts for the fact that our model is not exact, and prevents rare occurrences not described by the model from overwhelming the fit. $W$ is chosen to be 1, so ideally $W \approx \frac{\tilde{\chi}^2}{N} \approx 1$ \cite{Hauschild2001}.

We use the Matlab implementation of Differential Evolution \cite{DE_matlab} to optimize this function, with the following constraints:

\begin{itemize}
	\item The density of each emitter family must be large enough to have at least one defect in the sample area, but not so large that there is more than one defect per square Gaussian blur.
	\item The brightness of each emitter family must be larger than zero, but less than the maximum brightness of a single pixel in the sample minus the background parameter.
	\item The density of each each emitter family must be larger than the density of pixels of the family's average brightness expected due to the Poisson background
	\item The width of each emitter family must be larger than the Poisson width expected due to the family's brightness, but not larger than the family's brightness itself (otherwise it would predict negative brightness defects)
\end{itemize}

In addition, to decrease the parameter space volume (which increases the density of points sampled by the differential evolution algorithm), we restricted the parameters to be within an order of magnitude either way of the initial guess parameters (for emitter parameters) or within a factor of 2 for the background parameter (which is more tightly constrained). This is large enough to capture the behaviors we are looking for while still allowing for repeatable and accurate convergence of the optimization algorithm. Further rounds of convergence with even smaller windows were used to find better parameters, with the best parameter set (according to the Akaike Information Criteria, see below) always kept.

\subsection{Parameter Uncertainties}
To estimate the uncertainty, we assume that, near the minima, the chi-squared function looks parabolic:
\begin{equation}
\tilde{\chi}^2(\mathbf{x}) \approx \tilde{\chi}^2(\mathbf{x_0}) + \frac{1}{2}(\mathbf{x}-\mathbf{x_0})^TH(\mathbf{x_0})(\mathbf{x}-\mathbf{x_0})\,,
\end{equation}
where $\mathbf{x}$ is the vector of parameters, $\mathbf{x_0}$ is the argument of the minimum chi-squared, and $H(\mathbf{x_0})$ is the Hessian at $\mathbf{x_0}$.
$H(\mathbf{x_0})$ is calculated numerically using the DERIVEST suite. In cases where $H(\mathbf{x_0})$ is not positive definite, which is possible due to shallow or insufficiently converged minima, negative eigenvalues are corrected by finding the distance along that eigenvector necessary to produce a change in $\chi^2$ of 1.
Using the formula for log likelihood \cite{Hauschild2001},
\begin{equation}
log(\mathcal{L}) = -\frac{\tilde{\chi^2}}{2} + \mathrm{const}\,,
\end{equation}
where $\mathcal{L}$ is the likelihood of the estimated parameters, we estimate the standard errors to be
\begin{equation}
\epsilon_i \approx \sqrt{(H(\mathbf{x_0})/2)^{-1}_{ii}}\,.
\end{equation}
To get 95\% confidence intervals, we multiply the standard errors by 1.96.

\subsection{Selecting number of Defect Families}
To select the optimal number of families, we need a ``goodness of fit" metric which allows us to compare fits with different numbers of parameters.
Here we use the Akaike Information Criteria (AIC), defined by
\begin{equation}
AIC = \tilde{\chi^2}(\mathbf{x_0}) + 2*\mathrm{length}(\mathbf{x}) + \mathrm{const}\,,
\end{equation}
where a lower AIC is considered to represent a better fit.
Note that we again use the expression for log likelihood in terms of the modified Neyman chi-squared parameter.
We start with zero emitter families, and produce optimal fits for increasing number of families until the AIC begins to increase. We then take the best AIC and assume that represents the ``best" fit with the optimal number of emitter families.

\subsection{Testing the Model}
\begin{figure}[t]
\centering
\includegraphics[width=0.75\columnwidth]{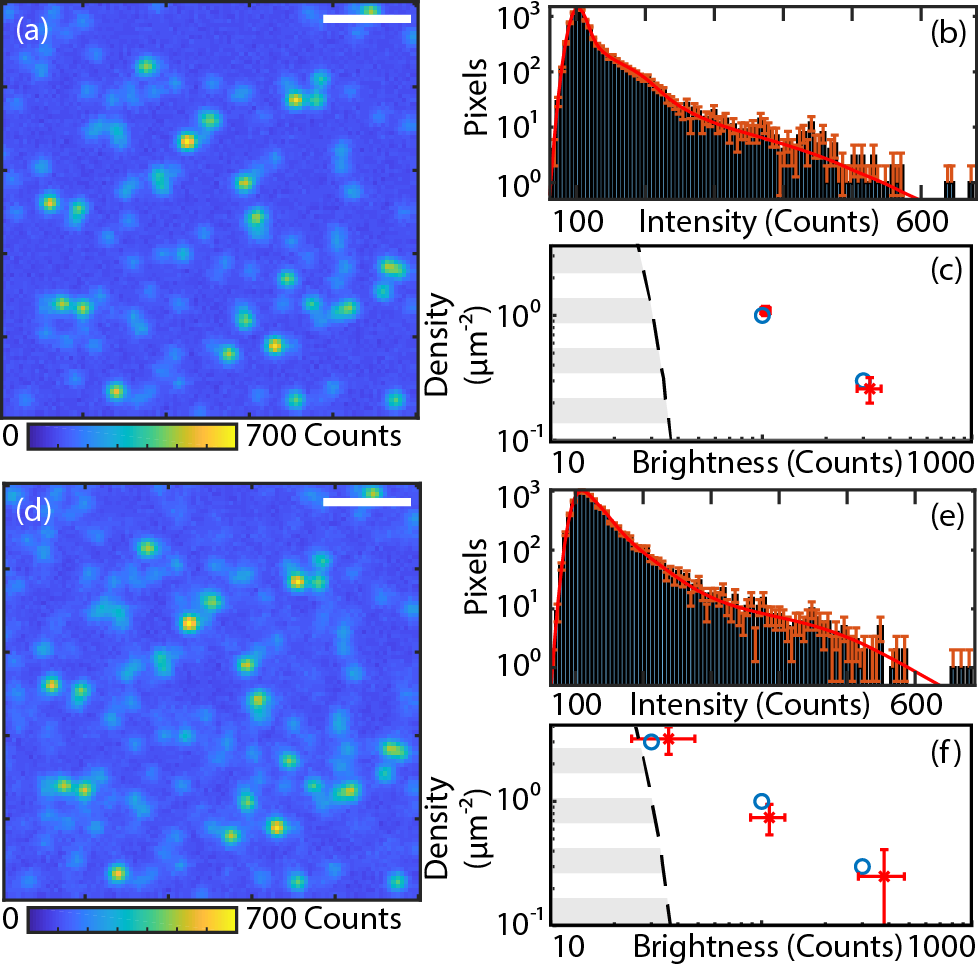}
\caption{\label{testing_multiple_families}a,d) Simulations similar to the one presented in FIG.2(d) in the main text, but with two and three emitter families, respectively. (b-c, e-f) Corresponding pixel intensity histograms and emitter family parameter plots. The emitter family parameters found by fitting agree with the underlying simulation values, shown by blue circles. Scale bars in (a,d) represents \SI{2}{\micro\meter}. Error bars in (c,f) represent \SI{95}{\percent} confidence intervals.}
\end{figure}
To test our model and fitting procedure, we apply it to simulated datasets with known parameters.
These simulations mirror the assumptions in our analysis - namely, emitters with brightnesses drawn from multiple normal distributions are placed randomly onto a Poissonian background with fixed densities.
Throughout our experimental datasets, most backgrounds seem to be around 100 counts with little variation between samples (prior to annealing).
We therefore fix our background to be $\lambda = \SI{100}{counts}$.
This sets a noise floor of $\sqrt{\lambda} = \SI{10}{counts}$, which in turn sets a scale for the brightness of emitters.
The result of simulating one emitter family, well above this scale, is shown in the main text in Fig. 2(d-f).
As shown there, the density and brightness of emitters found by the fitting procedure agrees with the true underlying parameters used in the simulation (as determined by the 95\% confidence interval of the fit).
In addition, the best-fit background of \SI{100.11 \pm 0.39}{counts} agrees with the simulated background parameter of \SI{100}{counts}.

\Cref{testing_multiple_families} presents similar simulations with two and three families.
\Cref{testing_multiple_families}(a) shows a simulation that adds an additional emitter family that is brighter but less dense than the first one.
Again, the fitted family parameters shown in \cref{testing_multiple_families}(c) are in agreement with the underlying simulation values, and the true background of \SI{100}{counts} is within the uncertainty range of the fitted background, \SI{99.75 \pm 0.28}{counts}.

\Cref{testing_multiple_families}(d) shows a simulation that further adds a dim, very dense emitter family that is difficult to distinguish by eye.
And again, the fitted family parameters shown in \cref{testing_multiple_families}(e) are in agreement with the underlying simulation values, and the true background of \SI{100}{counts} is within the uncertainty range of the fitted background, \SI{100.6 \pm 1.8}{counts}.
We note, however, that the fits are less accurate and the confidence intervals have increased, as expected due to the larger number of parameters.
However, this shows that the model is able to capture multiple families accurately.

\section{\label{sec:Samples}Samples}
\begin{table*}[h]
\begin{tabular}{|l|l|l|l|}
\hline
Region & Flake Thickness     & 1st Treatment                    & 2nd Treatment \\
\hline
A1     & 215nm               & \SI{2e16}{e\textsuperscript{-}\per\square{\centi\meter}} 3keV irradiation & 30 min \SI{850}{\celsius} Ar Anneal \\
\hline
A2     & 240nm               & \SI{2e16}{e\textsuperscript{-}\per\square{\centi\meter}} 3keV irradiation & 30 min \SI{850}{\celsius} Ar Anneal \\
\hline
B1     & 390nm               & \SI{2e17}{e\textsuperscript{-}\per\square{\centi\meter}} 3keV irradiation  & 30 min \SI{850}{\celsius} Ar Anneal \\
\hline
B2     & 250-350nm           & \SI{2e17}{e\textsuperscript{-}\per\square{\centi\meter}} 3keV irradiation  & 30 min \SI{850}{\celsius} Ar Anneal \\
\hline
C1     & 630nm               & (Ambient chamber conditions)     & 30 min \SI{850}{\celsius} Ar Anneal \\
\hline
C2     & 60nm                & (Ambient chamber conditions)     & 30 min \SI{850}{\celsius} Ar Anneal \\
\hline
C3     & \textgreater300nm & (Ambient chamber conditions)     & 30 min \SI{850}{\celsius} Ar Anneal \\
\hline
D1\textsuperscript{\textdagger}     &    *                & 30 min \SI{850}{\celsius} Ar Anneal            & \SI{4e15}{e\textsuperscript{-}\per\square{\centi\meter}} 3keV irradiation \\
\hline
D2     &    *                & 30 min \SI{850}{\celsius} Ar Anneal            & \SI{6e15}{e\textsuperscript{-}\per\square{\centi\meter}} 3keV irradiation \\
\hline
D3     &    *                & 30 min \SI{850}{\celsius} Ar Anneal            & \SI{1e16}{e\textsuperscript{-}\per\square{\centi\meter}} 3keV irradiation \\
\hline
E1     & 200-250nm           & \SI{2e16}{e\textsuperscript{-}\per\square{\centi\meter}} of 5keV irradiation & 30 min \SI{850}{\celsius} Ar Anneal \\
\hline 
\end{tabular}
\caption{\label{regions_all}A table of the regions studied and the treatments used.}
* Thickness information is not available for these regions.
\textsuperscript{\textdagger} This region underwent an additional 30 min \SI{850}{\celsius} Ar Anneal after being irradiated.
\end{table*}

\cref{regions_all} shows all regions for which data was taken, including the thickness of the flakes near the suspended regions. Regions B1 and B2 are on the same flake. Regions D1-D3 were exfoliated from a different bulk crystal and underwent an $O_2$ plasma clean prior to initial imaging.

\section{Additional Emitter Distributions}

\begin{figure}[t]
    \centering
    \begin{minipage}{0.45\columnwidth}
        \centering
        \includegraphics[width=\columnwidth]{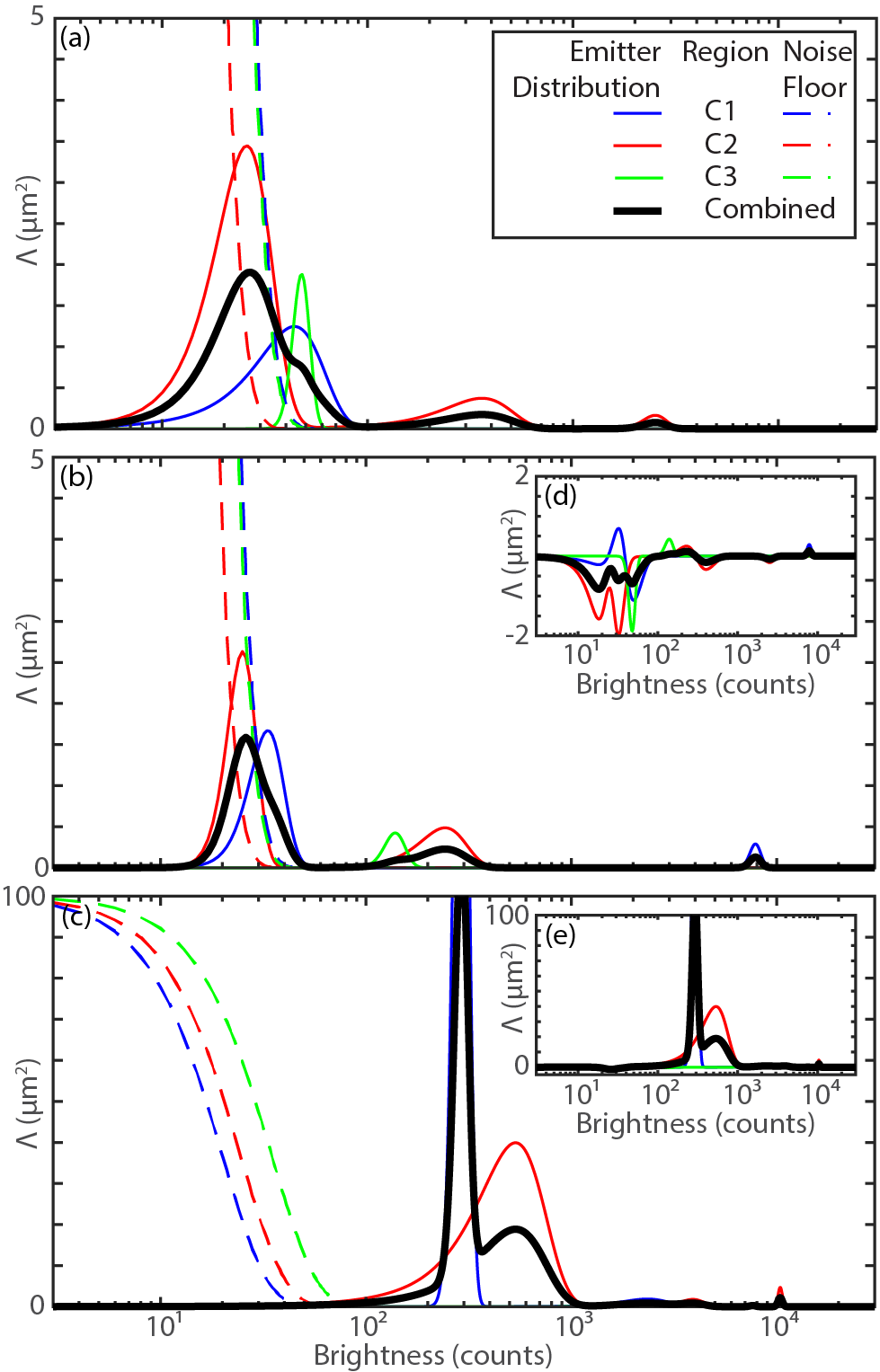}
        \caption{\label{control_distrubtion_plots}Emitter distributions for regions which received indirect irradiation prior to annealing; distributions for each region a) before treatment, b) after irradiation, and c) after annealing are shown. Changes are shown for the d) pre-treament to post-irradiation and e) post-irradiation to post-annealing distributions.}
    \end{minipage}\hfill
    \begin{minipage}{0.45\columnwidth}
        \centering
        \includegraphics[width=\columnwidth]{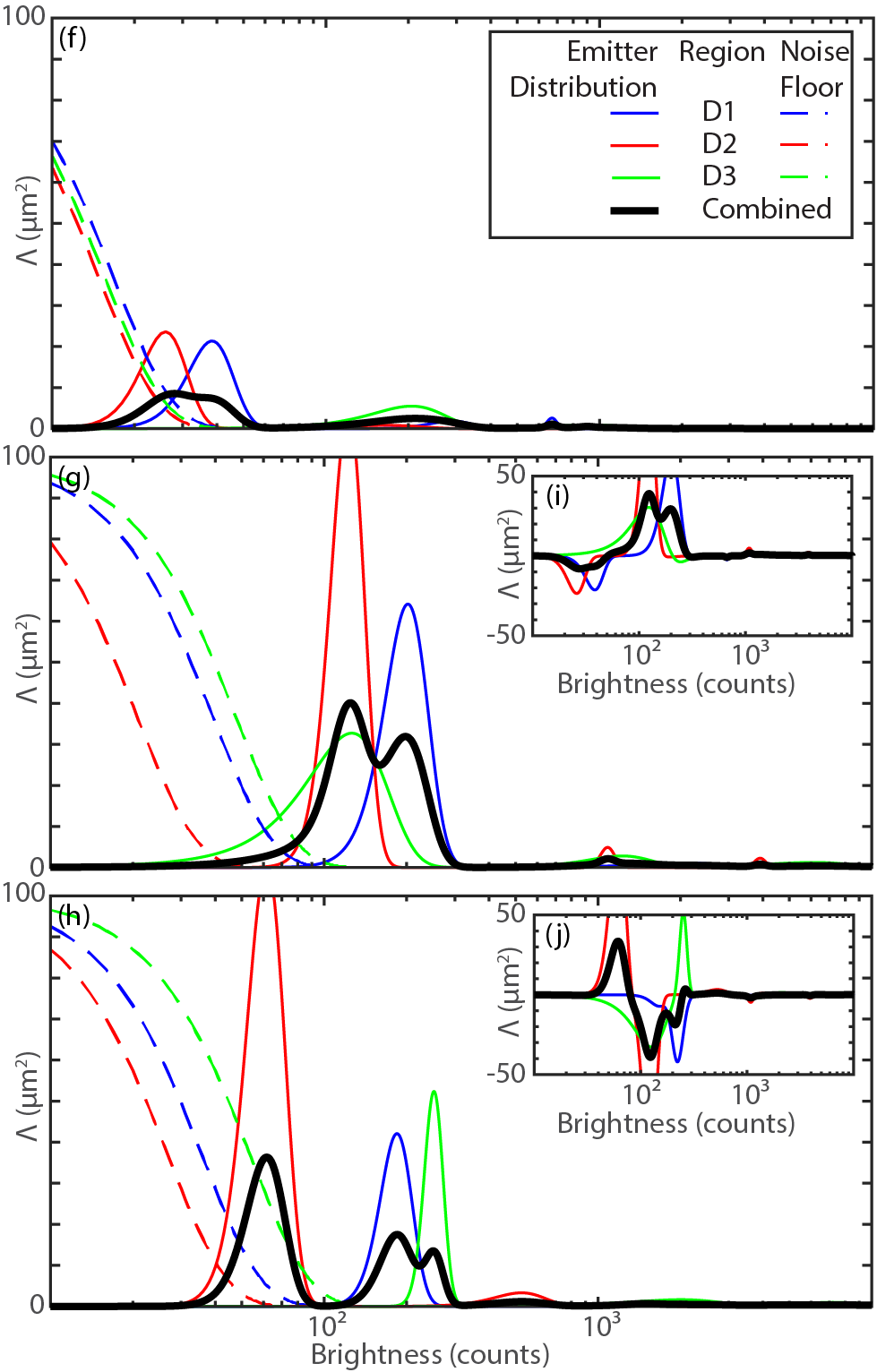}
        \caption{\label{annealed_first_distrubtion_plots}Emitter distributions for regions which received annealing prior to direct irradiation; distributions for each region a) before treatment, b) after annealing, and c) after irradiation are shown. Changes are shown for the d) pre-treament to post-annealing and e) post-annealing to post-irradiation distributions.}
    \end{minipage}
\end{figure}


\Cref{control_distrubtion_plots} and shows emitter distributions similar to those shown in Fig. 4 in the main text, but for regions which received only indirect irradiation followed by annealing.
The pre-treatment emitter distribution in \cref{control_distrubtion_plots}(a) is similar to that presented in Fig. 4(a) of the main text, dominated by a peak around \SI{40}{counts}, with additional smaller peaks at higher brightnesses.
After receiving only indirect exposure, the distribution in \cref{control_distrubtion_plots}(b) shows the same peaks but with decreased densities, which we attribute to photobleaching.
In addition, a small peak at high brightness appears, which is attributed to irradiation from stray ions caught in the accelerating voltage.
After annealing, a huge density of emitters is found between \SI{100} and \SI{1000}{counts}, which we attribute to emitter present before annealing but below the noise floor becoming much brighter.
In addition, smaller peaks are found above \SI{1000}{counts}, which we attribute to brightening of emitters already visible in the sample.

\cref{annealed_first_distrubtion_plots} again shows emitter distribution, but for regions which were annealed prior to being irradiated.
\cref{annealed_first_distrubtion_plots}(a) shows a pre-treatment emitter distribution which much denser broader than those in \cref{control_distrubtion_plots}(a) or Fig. 4(a), which we attribute to the flakes being exfoliated from a different bulk crystal.
After annealing, the distributions in \cref{annealed_first_distrubtion_plots}(b) become much brighter and denser.
In addition, the background of two of the three regions increased significantly.
We attribute these effects to both already visible emitters becoming brighter, as well as emitters from below the noise floor rising above it.
After irradiation, the distributions in \cref{annealed_first_distrubtion_plots}(c) are both slightly dimmer and slightly less dense than those in \cref{annealed_first_distrubtion_plots}(b).
We attribute this again to photobleaching, and note that, based on the changes in Fig. 4(b) of the main text, the effects of the low dosage of irradiation used would not be visible on this scale.

\section{Raw Data}
We also include (as an ancillary document) all of the raw data taken in the study, as well as the parameter plots from the corresponding fits. For each stage of the treatment, the PL map is shown with a \SI{1}{\micro\meter} scale bar, and the suspended region is outlined in red. For regions where supported photoluminescence near the edge of the suspended region appear to bleed into the suspended region, a buffer uniformly shrinks the suspended region (so as to not introduce any sampling bias). Along with the PL maps, pixel histograms and the fit resulting from the described fitting procedure are presented.

\bibliography{supplement}